# The influence of solvent representation on nuclear shielding calculations of protonation states of small biological molecules

Christina C. Roggatz[1], Mark Lorch[1] and David M. Benoit[1,2]

[1]Chemistry – School of Mathematics and Physical Sciences, University of Hull, Cottingham Road, Hull, HU6 7RX, UK;
[2]E.A. Milne Centre for Astrophysics & G.W. Gray Centre for Advanced Materials, University of Hull, Cottingham Road, Hull, HU6 7RX, UK

E-mail: roggatz@outlook.com; D.Benoit@hull.ac.uk

## Abstract

In this study, we assess the influence of solvation on the accuracy and reliability of nuclear shielding calculations for amino acids in comparison to experimental data. We focus particularly on the performance of solvation methods for different protonation states, as biological molecules occur almost exclusively in aqueous solution and are subject to protonation with pH. We identify significant shortcomings of current implicit solvent models and present a hybrid solvation approach that improves agreement with experimental data by taking into account the presence of direct interactions between amino acid protonation state and water molecules.

## 1. Introduction

Amino acids are the central building blocks of life and involved in many biochemical processes. The nuclear magnetic resonance (NMR) characteristics of these small biomolecules have been found to be crucial for the assessment of ligand binding, in pharmacological studies, during the assignment of chemical shifts in complex systems and for conformational studies.[1] The chemical shifts of amino acids and very small peptides can be used, for example, to serve as "random coil" baseline for protein structure investigations.[2] Any systematic deviations from these "baseline shifts", which depend primarily on the identity of the amino acid side chain and conformation, can be used to identify secondary structures such as α-helices or β-strands.[2] Chemical shifts not only provide insight into the molecular conformation of the close surrounding of a nucleus but also indicate changes in their chemical environment.[2,3] This makes shifts ideal parameters for the investigation of molecular processes. Changes in chemical shift can give information on neighbouring groups or amino acid sequences in a peptide or protein.[4] Amino acid side chains are also key components during catalysis in enzymes or in binding pockets of receptors and changes in their chemical shifts can help to analyse reaction pathways and to indicate substrate binding.[5] However, NMR spectroscopy has its limitations when it comes to investigations of the underlying mechanisms at the atomistic level.[6] To counteract these limitations, studies often make use of computational simulations and modelling to gain more detailed insights (see for example Han *et al.*[7]). Amino acids are therefore also increasingly studied with computational methods in attempts to optimise computer simulation programs and protocols.

In order to validate computational simulations, it is crucial to compare results obtained through modelling to the available experimental data. However, this requires a realistic calculation of chemical shift values. In practice, chemical shifts are measured as change of resonance frequency of a nucleus relative to a given standard, for example the same type of nucleus in tetramethylsilane (TMS).[8] The resonance frequency of a nucleus, and therefore its chemical shift, directly depends on the magnetic field that the nucleus experiences. This differs for each nucleus due to shielding or deshielding effects caused by electrons surrounding the nucleus, which can be described by a nuclear shielding constant $\sigma_{nuc}$. This nuclear shielding constant can be calculated at quantum chemical level (see Mulder & Filatov[2] for an overview). The chemical shift of a given nucleus ($\delta$) can then be derived from the calculated shielding constant $\sigma_{nuc}$ and the shielding constant of the same nucleus type in the standard reference compound ($\sigma_{ref}$) used in the NMR experiments (e.g. TMS):[9]

$$\delta = (\sigma_{ref} - \sigma_{nuc}) \tag{1}$$

The reference shielding value $\sigma_{ref}$ can be obtained by a separate calculation performed for the reference compound. As can be seen in Eqn. (1), the chemical shift of a nucleus ($\delta$) is related to the negative of its nuclear shielding value ($\sigma_{nuc}$) due to the historical custom of a reversed frequency scale. This means nuclei that are more shielded than the reference nucleus have lower chemical shifts and those less shielded have higher chemical shifts. As biomolecules appear almost exclusively in aqueous solution, the contribution of solvation ($\sigma_{solvent}$) is a major factor impacting on the value of the shielding of a nucleus ($\sigma_{nuc}$). In addition, there are several interactions between the solvent and solute that play critical roles in stabilising conformations and mediating molecular processes.[10] Buckingham *et al.* identified four different contributions to the shielding effect of solvent $\sigma_{solvent}$:[11]

$$\sigma_{solvent} = \sigma_b + \sigma_a + \sigma_E + \sigma_w \tag{2}$$

which is composed of a long-range bulk diamagnetic susceptibility effect ($\sigma_b$), the anisotropy in the molecular susceptibility of the solvent molecules close to the solute ($\sigma_a$), a polar effect ($\sigma_E$) and Van der Waals forces between solute and solvent molecules ($\sigma_w$).[11] While $\sigma_a$ is particularly important for solvents with large $\pi$ systems, the polar effect $\sigma_E$ and van der Waals forces $\sigma_w$ can be assumed to dominate in aqueous solutions[12]. The polar effect ($\sigma_E$) is caused by the charge distribution in the solvent molecules leading to the formation of an electric field that perturbs the electronic structure of the solute.[11] This clearly has a strong effect on the magnetic shielding at the nuclei. Interactions through hydrogen bonds can be seen as a special manifestation of the polar effect ($\sigma_E$) and have been shown to significantly affect $^1$H chemical shift values of those protons directly involved in the H-bond.[13] Van der Waals forces between solute and solvent molecules ($\sigma_w$) can also contribute significantly to the overall solvent shift.

In order to appropriately model solvent interactions, a number of different solvation methods has been developed ranging from periodic molecular dynamic simulations of the solvent to implicit, explicit or hybrid models in *ab initio* and DFT calculations. The implicit solvation model averages the effects of all solvent molecules around the solute[14] and simplifies the interactions of the solute-solvent system by describing the solvent with a single dielectric constant.[15] The solute is placed into a cavity of a continuous polarisable medium (= solvent) and the interaction of the solute with the surrounding field is calculated at the cavity boundaries[15,14]. Amongst the implicit solvent models are the well-known polarised continuum model (PCM)[14,16] and the conductor-like screening model (COSMO)[17]. More refined models include the integral equation formalism (IEFPCM)[18], the COSMO-RS (realistic solvents)[19] and the reference interaction site model (RISM)[20]. These models are frequently used during NMR calculations. However, while implicit solvation remains a computationally inexpensive approach that captures the effect of bulk solvent ($\sigma_b$) well and is particularly suitable for large solutes, it often neglects local effects within the first solvent shell such as strong hydrogen bonds.[21] These local effects are better represented by explicit solvation, also referred to as microsolvation,[21] where individual solvent molecules are placed around the solute[22]. The number of solvent molecules can be varied but is kept as low as possible, typically one to ten,[22] to reduce computational costs. Solute molecules can also be specifically allocated to certain functional groups or atoms of the solute in larger systems. This allows to accurately represent short-range interactions between solute and solvent, such as $\sigma_E$ and $\sigma_W$.[21] But it also poses the risk of forming additional interactions with the solute or have dangling O–H bonds and lone pairs that would not be present experimentally.[21] Explicit solvation also is more computationally costly than the use of implicit solvation models. A combination of both models creates a hybrid solvation model, where a cluster of the solute and a small number of explicit solvent molecules is placed into the implicit dielectric field.[21] This model provides a promising cost-effective way of treating solvation in large solute-solvent systems with strong local effects. It has previously been called a combined discrete-continuum model, cluster-continuum model or implicit-explicit model.[23] For small biomolecules in aqueous solution it can be assumed that both the bulk solvent as well as local effects play a significant role in shielding the nuclei, but this has not been investigated systematically for amino acids to our knowledge.

Apart from causing significant solvent-solute interactions, water can give rise to a third factor influencing NMR properties: protonation. Protonation is an ubiquitous process in biology and the most common ionisation process in proteins.[24] Ionisation and proton transfer play significant roles in electrostatic interactions, ligand recognition, protein folding, enzyme catalysis, membrane potentials and the energetics of cells.[24] Depending on the pH of the aqueous solution different protonation states of the same amino acid can be present at the same time. Even within the human body, pH ranges from 1.5 in the stomach to 8 in the pancreas.[25] It is therefore essential to investigate different protonation states of biological molecules. However, the focus of research to date has been on the protonation states present at physiological pH, which refers to the well buffered blood pH of 7.4 ($\pm$ 0.05), and neglected other protonation states.

In this study, we investigate the effect of solvation on nuclear shielding calculations for amino acids. Our test set includes all protonation states of four different amino acids and we assess the influence of the implicit solvent method, the magnetic Hamiltonian and the hybrid solvation model performance on the accuracy and reliability of their computed nuclear shielding.

## 2. Methods

### 2.1 Amino acid test set

For this study, we use a test set of four amino acids: glycine, L-alanine, L-cysteine and L-serine. For each amino acid, we investigate all protonation states present in a pH range of 0 – 14 (Fig. 1). For the calculations of the protonated states of glycine, L-serine and L-cysteine, the lowest energy conformation according to Balabin[26] and Noguera *et al.*[27] were rebuilt using AVOGADRO (Version 1.1.0)[28] and optimised using the UFF force field[29] and a steepest descent algorithm. The xyz coordinates of each preliminarily optimised molecule were then used as starting conformation for further geometry optimisation. The zwitterionic and deprotonated forms were based on the structure of the neutral form and hydrogen atoms were added or removed as necessary. For L-alanine, the optimised neutral conformation was taken from Godfrey *et al.*[30] and de-/protonated forms were obtained by subtraction or addition of a proton to the neutral structure.

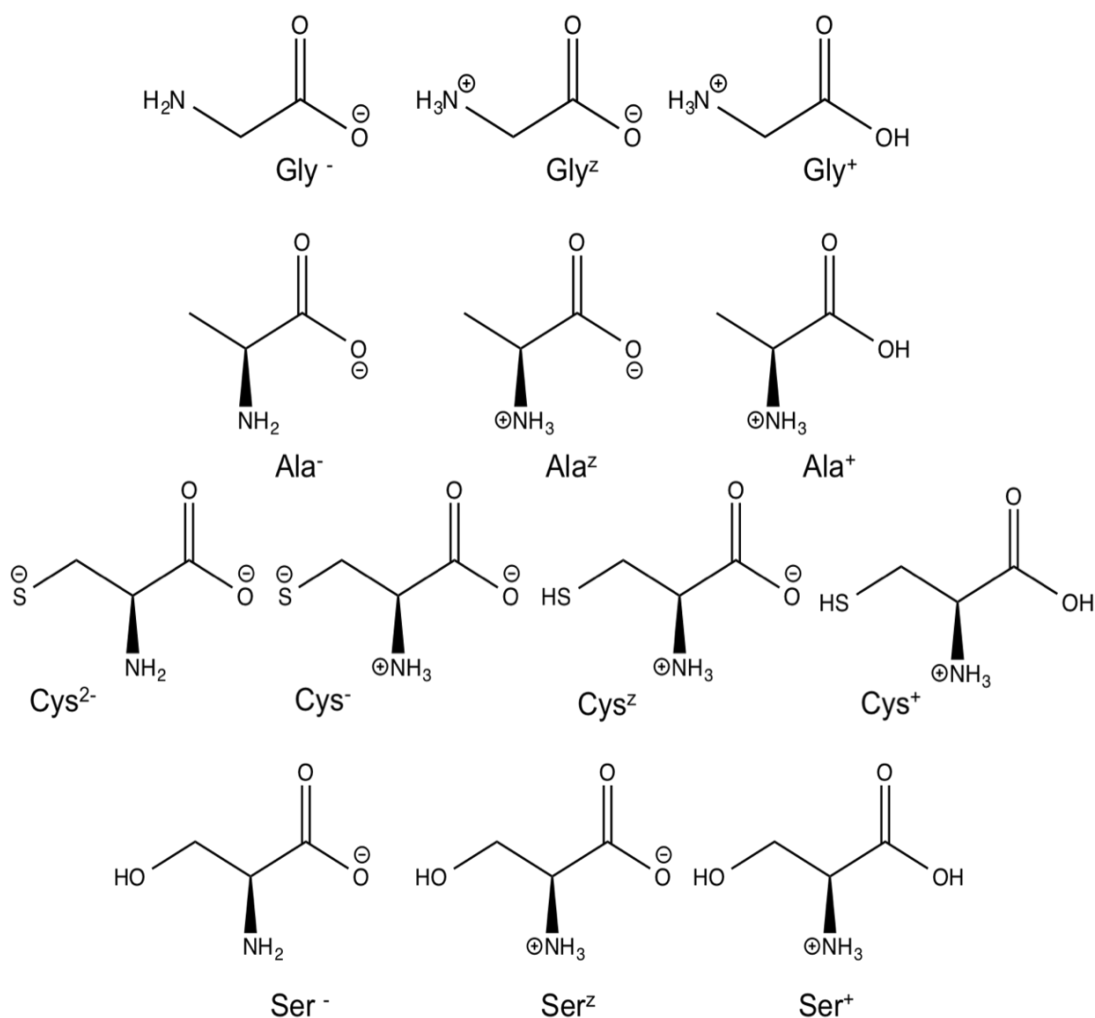

*Fig. 1:* Protonation states of the non-polar amino acids glycine and L-alanine and the polar amino acids L-cysteine and L-serine used in the test set.

## 2.2 Computations

Geometry optimisations were performed using the PBE0 exchange correlation functional[31] with a pc-2 basis set[32,33,34]. The RIJ-COSX approximation[35] was used with a def2-TZVPP/J auxiliary basis set[36] including D3 dispersion corrections following Grimme [37,38]. Shielding values of $^1$H and $^{13}$C nuclei were calculated at the PBE0/aug-pc-2 level of theory, using the RIJ-COSX approximation with a def2-TZVPP/J auxiliary basis set and the individual gauge for localised orbitals method (IGLO)[39]. All calculations were run in ORCA (Version 3.0.0[40]).

The DFT functional PBE0 has been chosen as it is known to perform very well for NMR calculations,[9,41] especially for nuclei with electrostatic interactions and in biological systems[42]. Single point energy calculations and geometry optimisations[42,43,44] as well as nuclear shielding calculations performed with PBE0[42,45,44] were previously found to be in good agreement with experimental data. PBE0 is therefore amongst the most popular functionals used in NMR calculations. However, for the basis set we opted against the widely used valence triple-$\zeta$ Pople basis sets[46,47] (see for example in Bachrach[48] and Baggioli[42]). Instead we chose the pc-$n$ and aug-pc-$n$ basis sets developed specifically for DFT calculations by Jensen[32] as they were reported to perform better than the Pople basis sets[49]. The aug-pc-$n$ set contains added diffuse functions, which substantially improve basis set convergence for molecular properties that depend on regions far from the nuclei, such as electric multipole moments and polarisabilities[49] and was therefore used for the NMR nuclear shielding calculations. In a comparative test aug-pc-2 was found to perform better than the aug-pcS-2 basis set[49] (see Supplementary Material T1), which has been developed specifically for chemical shielding calculations. The aug-pc-2 basis set showed a higher accuracy and reliability (correlation of fit) at a lower computational cost compared to aug-pcS-2 in our test for the amino acid test set with implicit solvation for both nuclei.

## 2.3 Experimental measurements of chemical shifts

NMR samples were prepared for all four different amino acids from their aminoacetic acid forms (powder, >99% purity level, 20mg, Sigma Ultra, Sigma-Aldrich, UK) in sodium phosphate buffer (10 mM, with 10% $D_2O$) at a 100 mM concentration and with TMS as internal standard. TMS was used to calibrate the experimental NMR spectra, as it is insensitive to the different pH conditions, and set at 0 ppm. For each amino acid samples of different pH were prepared in which the respective amino acid was mainly present in protonated, zwitterion, deprotonated or fully deprotonated form (Cys only) based on their literature $pK_a$ values[50] (Gly: pH 1.12, 6.51 and 11.99; Ala: pH 0.95, 6.20 and 12.17; Ser: pH 0.87, 5.35 and 12.41; Cys: pH 1.07, 4.89, 9.19 and 12.77).

NMR spectra were measured using a Bruker Advance II 500 MHz spectrometer at 298K. Proton chemical shifts were determined using the WATERGATE method[51] for water suppression and proton resonance was measured at 500 MHz. Proton chemical shifts were determined with a 1D sequence with water suppression using the 3-9-19 pulse sequence with gradients[51,52] and 64 scans $^{13}$C chemical shifts were measured with the 1D $^{13}$C sequence in decoupled mode. Spectra were processed using Topspin Version 2.0 (Bruker Instruments, Karlsruhe, Germany).

## 2.4 Comparison of computed nuclear shielding and experimental chemical shift values

Experimentally, only a single chemical shift could be obtained for protons of $CH_2$, methyl and amino groups. This is due to the fact that the exchange rate (rotation) of the protons in these groups is fast compared to the small difference in nuclear frequency and leads to averaging over the system.[53] To facilitate comparison to the measured data, the calculated nuclear shielding values of protons of these groups were therefore averaged.

Calculated shielding constants ($x$-axis) were plotted against experimental chemical shift values ($y$-axis) and a least-square linear curve fit was performed with IGOR pro (Version 6.02, WaveMetrics, Inc. 1988-2007) using the linear function:

$$\delta_{exp} = a + (b \cdot \sigma_{calc}) \quad (3)$$

where $a$ is the intercept with the $y$-axis and $b$ the slope of the line. The quality of the least-squares fit was assessed using the Pearsons correlation coefficient, $R^2$. This direct comparison using calculated nuclear shielding values helps to avoid offset errors[1]. It further allows a direct analysis of method accuracy without interference by error cancellation, which can be observed when chemical shift values are calculated as difference between a molecular and a standard nucleus (see Eqn. (3)). The slope of the correlation is expected to be close to $-1$ due to the inverse relationship between nuclear shielding and chemical shift values. The intercept with the $x$-axis should in theory represent the shielding value of the $^1$H and $^{13}$C nuclei for the reference substance TMS. Therefore substituting the $x$ value of the linear curve fit function with the shielding value calculated for TMS should give 0. Any deviation from 0 gives an indication of the error of the chosen calculation methods. In order to determine accuracy, the shielding constants of the $^1$H and $^{13}$C nuclei of TMS were calculated with the methods described above for the amino acid forms and with the respective solvation models. The calculated values for TMS were substituted into the linear curve fit function and the function was solved for $y$. The difference between the result of the linear function and the expected value of 0 ppm was stated as accuracy.

## 3. Results and Discussion

### 3.1 The importance of solvation: Gas phase vs. COSMO

We first optimised the geometries of the test set both gas phase and in implicit solvent (water) using the COSMO[19] approach implemented in ORCA. The nuclear shielding values of each geometry was then calculated as described in section 2.2. The results were plotted in comparison to experimental chemical shift values determined for the respective nuclei. Carbon nuclear shieldings ($^{13}$C) (Fig. 2A) were found to correlate better with experimental values by 0.05% when calculated with implicit COSMO solvation ($R^2 = 0.9990$) compared to gas phase. The accuracy of the nuclear shielding calculation was improved significantly by COSMO solvation, enhancing it by 2.30 ppm to an overall accuracy of $\pm$ 2.92 ppm.

For hydrogen nuclei ($^1$H, protons) (Fig. 2B) calculated nuclear shielding values were found to correlate better with experimental values by 4.88% when calculated with implicit solvation (COSMO) compared to gas phase. The overall correlation, however, was very weak with $R^2 = 0.1987$ for COSMO solvation. The accuracy of the proton nuclear shielding calculation was improved significantly by implicit solvation with COSMO, enhancing it by 2.21 ppm to an overall accuracy of $\pm$ 1.09 ppm.

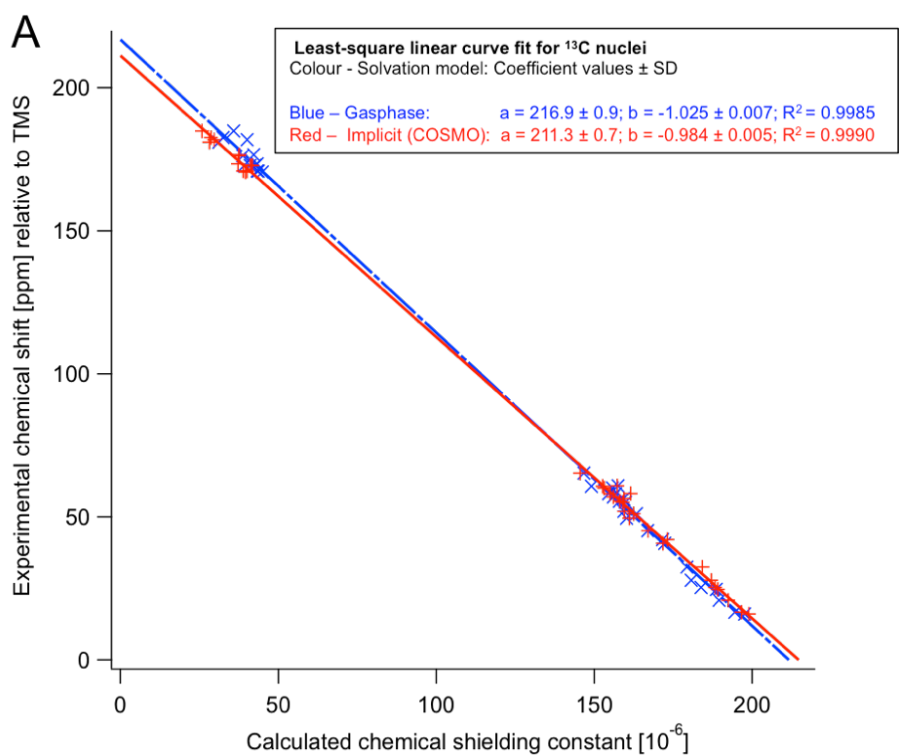

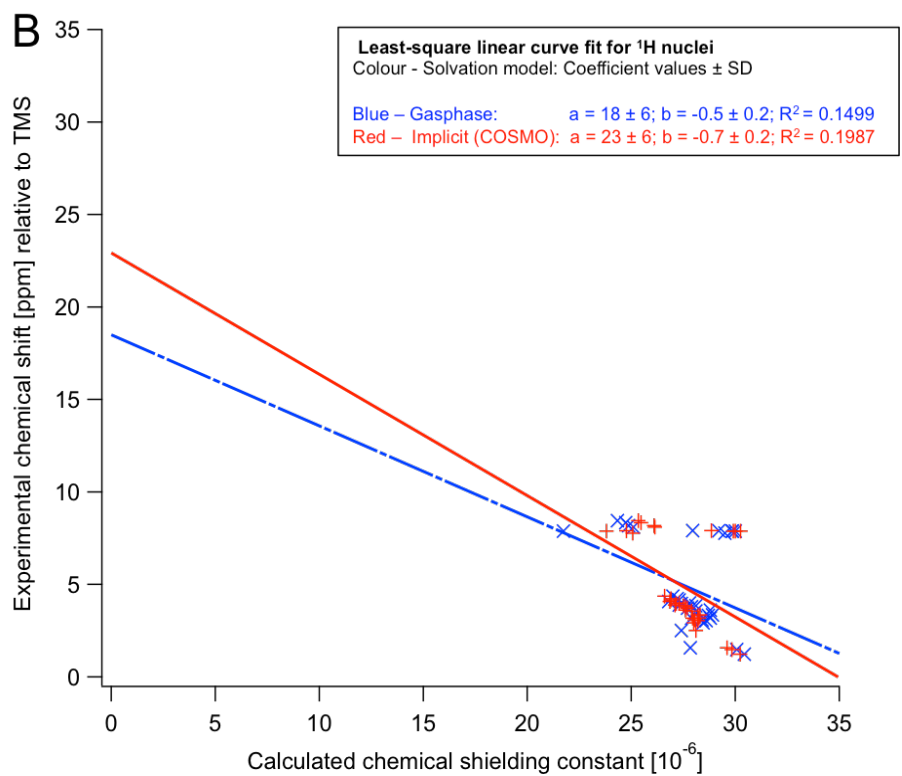

*Fig. 2: Correlation between nuclear shielding values calculated in gas phase or COSMO and experimental chemical shift values for $^{13}C$ (A) and $^{1}H$ (B) nuclei. For both nuclei types, calculations were performed in gas phase (blue, intermitted line) and implicit solvation (COSMO) (red, continuous line).*

These results clearly show that solvation matters and significantly improves nuclear shielding calculations to better match experimental values. This is consistent with recommendations given for NMR calculations in commonly used program suites like ORCA and Gaussian.[40,54] For $^{13}$C nuclei, the correlation between calculated and experimental values is very high and is only improved slightly by the implicit solvation, while the accuracy is improved by 56%. The slope of the linear regression is close to $-1 \pm 0.05$, indicating a well-performing method.[9] In contrast, the correlation for $^{1}$H nuclei is very poor, even when using implicit solvation (COSMO). The accuracy of proton nuclear shielding calculations, however, was improved by 33% upon solvation. The slope of the linear regression differs significantly from the expected values of around $-1$, indicating a high systematic error with the chosen method.

These findings contrast with the statements of Sousa *et al.*,[55] who concluded that conformation of amino acids optimised in gas phase present a reasonable alternative to those optimised with implicit solvation. Our results indicate that this is not the case when the conformations are used for calculations of NMR parameters. During our optimisations in gas phase, the amino acid conformers underwent proton relocation in zwitterionic protonation states. The accurate representation of amino acids in their natural protonation states, however, is critical to gain biologically meaningful results that can be used in future simulations. The inclusion of solvation during the geometry optimisation process is therefore essential for realistic nuclear shielding calculations.

## 3.2 Other implicit solvent models and the influence of the magnetic Hamiltonian

After establishing the importance of solvation for one methodology, we wanted to ensure that our findings are also consistent across different implicit solvation models, such as the widely used IEFPCM[18] implemented in Gaussian. We therefore calculated the nuclear shielding values of $^{1}$H and $^{13}$C with IEFPCM at the same PBE0/aug-pc-2 level of theory in Gaussian (Gaussian 09, Revision B.01)[56] using the optimised geometries obtained from the ORCA calculations. While ORCA only implements the individual gauge for localised orbitals (IGLO) method[39] for the magnetic Hamiltonian, the method to circumvent the so-called gauge problem implemented in Gaussian is the gauge-inducing atomic orbitals (GIAO) approach[57,58]. In order to obtain nuclear shielding constants independent from the magnetic field, both methods permit to mathematically cancel out the gauge origin by employing specific phase factors at atomic orbital (GIAO) or molecular orbital (IGLO) level.

To investigate the effects of the IEFPCM model, we calculated the nuclear shielding constants in gas phase and with this implicit solvent model[14,16]. The results are shown in Fig. 3 for both $^{13}$C (3A) and $^{1}$H (3B). The effect of using GIAO instead of IGLO as magnetic Hamiltonian was evaluated by comparing the gas phase results obtained in this section and those of section 3.1 and is discussed below.

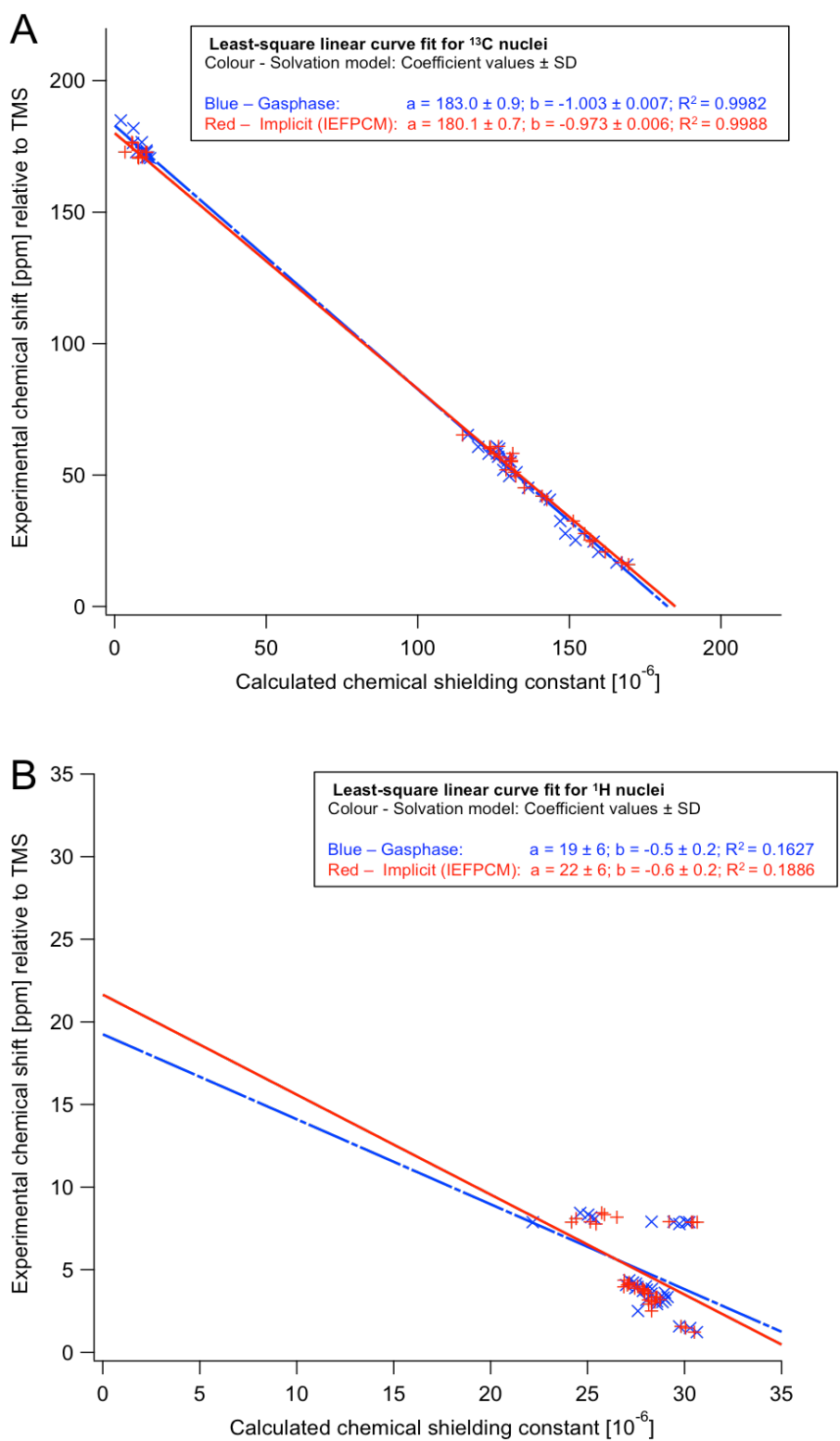

*Fig. 3: Correlation between nuclear shielding values calculated in gas phase or IEFPCM and experimental chemical shift values} for $^{13}$C (A) and $^{1}$H (B) nuclei. For both nuclei types, calculations were performed in gas phase (blue, intermitted line) and implicit solvation (IEFPCM) (red, continuous line).*

Carbon nuclear shieldings (see Fig. 3A) were found to correlate better with experimental values by 0.06% when calculated with IEFPCM solvation ($R^2 = 0.9988$) compared to gas phase. The accuracy of the nuclear shielding calculation was improved significantly by IEFPCM solvation, enhancing it by 2.06 ppm to an overall accuracy of $\pm$ 2.80 ppm. Proton shieldings (Fig. 3B) were found to correlate better with experimental values by 2.59% when calculated with IEFPCM compared to gas phase. The overall correlation, however, was again very weak with $R^2 = 0.1886$ for IEFPCM solvation. The accuracy of the nuclear shielding calculation was improved by IEFPCM solvation, enhancing it by 0.16 ppm to an overall accuracy of $\pm$ 3.06 ppm.

The results for the two different implicit solvent methods are very similar for $^{13}$C nuclei (Fig. 2A & Fig. 3A) when comparing the respective gas phase to implicit solvation results. For both solvation methods, slight changes to the correlation but clear improvements of the calculation accuracy are observed. For proton shieldings, both implicit solvent models show very poor correlation and when using IEFPCM solvation the accuracy is only improved slightly while in the COSMO calculation it is improved by a third.

The influence of the two different magnetic Hamiltonians, IGLO and GIAO, can be compared by looking at the gas phase results in Figs. 2 and 3. For both Hamiltonians the correlation between calculated and experimental data was very good for carbon nuclei and similarly poor for hydrogen nuclei. The accuracies for carbon differed by 0.3 ppm (< 6%) and for protons by less than 0.1 ppm (< 2.5%) between the two magnetic Hamiltonians. These differences are significantly smaller than the influence of implicit solvation. The choice of magnetic Hamiltonian is therefore of minor importance, which matches results of Facelli who reported calculations with sufficiently large basis sets to converge to the same nuclear shielding values with both methods[8].

In comparison to previously reported accuracies, which can be as low as $\pm$ 10 ppm for $^{13}$C [9], the accuracy of our approach with implicit solvation is relatively high with $\pm$ 2.92 ppm (COSMO) and $\pm$ 2.80 ppm (IEFPCM). For carbon nuclei, accuracies of around $\pm$ 5.00 ppm are frequently reported, see for example Sefzik et al. [59] or Kupka et al. [60].

Common accuracies for proton shielding calculations are in the region of $\pm$ 0.4 ppm[9]. This is significantly better than $\pm$ 1.09 ppm (COSMO) and $\pm$ 3.06 ppm (IEFPCM) obtained in our calculations. Frank et al. stated that the main reason for errors in proton chemical shifts is the neglect of explicit solvent molecules[1]. Proton chemical shifts are known to be highly influenced by their environment and H-bonds with solvent molecules can significantly alter their shielding by giving rise to the polar shielding effect ($\sigma_E$) (see Eqn. (2)). We therefore went on to test the effect of additional explicit solvent molecules on the nuclear shielding calculation.

### 3.3 Hybrid models

In order to improve the performance of our model for protons, we investigated how adding explicit water molecules affects the correlation and agreement of nuclear shielding constant calculations with experimental data. For this purpose, we created two types of hybrid solvation models by adding either one water molecule per ionisable functional group of the solute (low level hybrid, LH) or one water molecule per ionisable proton of the solute (high level hybrid, HH) (see Fig. 4). For the low hybrid model, water molecules were added directly next to the functional groups of the solute using AVOGADRO and the geometry optimisation and nuclear shielding constant calculations were performed as described in section 2.2 with COSMO as implicit solvation. For the high hybrid model, a two step process was applied to ensure the saturation of all protons with a respective water molecule. First, water molecules were added directly next to each ionisable proton of the solute using AVOGADRO, followed by a geometry optimisation at HF-3C level of theory[61]. This method is specifically developed for the fast computation of structures and non-covalent interactions in large molecular systems[61]. Optimised geometries were checked for saturation and, if required, more water molecules were added followed by another geometry optimisation at HF-3C level of theory until all ionisable protons were saturated with an H-bond from a water molecule. Then a full geometry optimisation and nuclear shielding constant calculation were performed as described in section 2.2 with COSMO as implicit solvation. For the calculation of TMS with the high hybrid solvation model, a symmetry approach was used and the nuclear shieldings for only one $^{13}$C and three $^{1}$H nuclei were calculated upon saturation with hydrogen bonds to explicit water molecules.

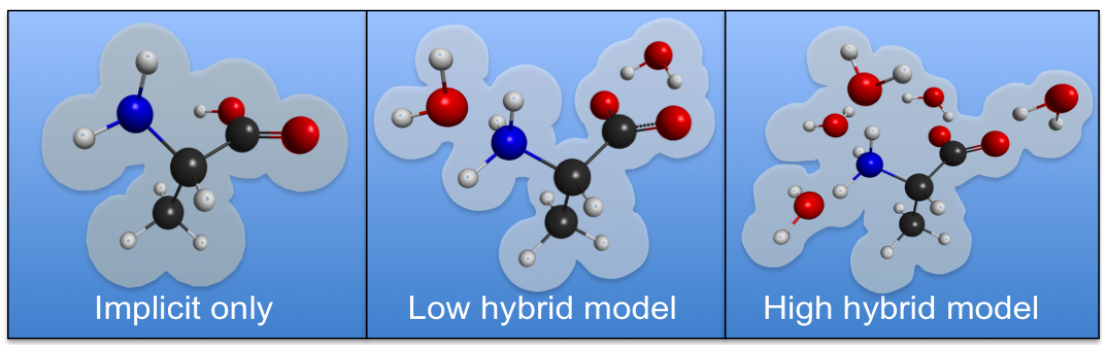

*Fig. 4: Schematic presentation of the different solvation models for L-alanine: implicit (left), low hybrid (LH, center) with one water molecule per ionisable group and high hybrid (HH, right) with one water molecule per ionisable proton.*

For carbon nuclei (Fig. 5A) the correlation with experimental values was only very marginally affected when calculated with low hybrid solvation ($R^2 = 0.9994$) or high hybrid solvation ($R^2 = 0.9989$) compared to implicit solvation only ($R^2 = 0.9990$). The accuracy of the nuclear shielding calculation was clearly higher ($\pm$ 2.48 ppm) for the low hybrid solvation than for calculations with implicit solvation only. For high hybrid solvation model the accuracy however was much lower ($\pm$ 5.5 ppm).
For protons (Fig. 5B), the correlation to experimental data could be significantly improved by 13.03% for low hybrid and by 25.84% for high hybrid solvation compared to implicit solvation only. In contrast, the accuracy of the nuclear shielding calculations with hybrid models was lower than with the implicit model. In all three solvation models a group of clear outliers from the expected linear correlation could be observed.

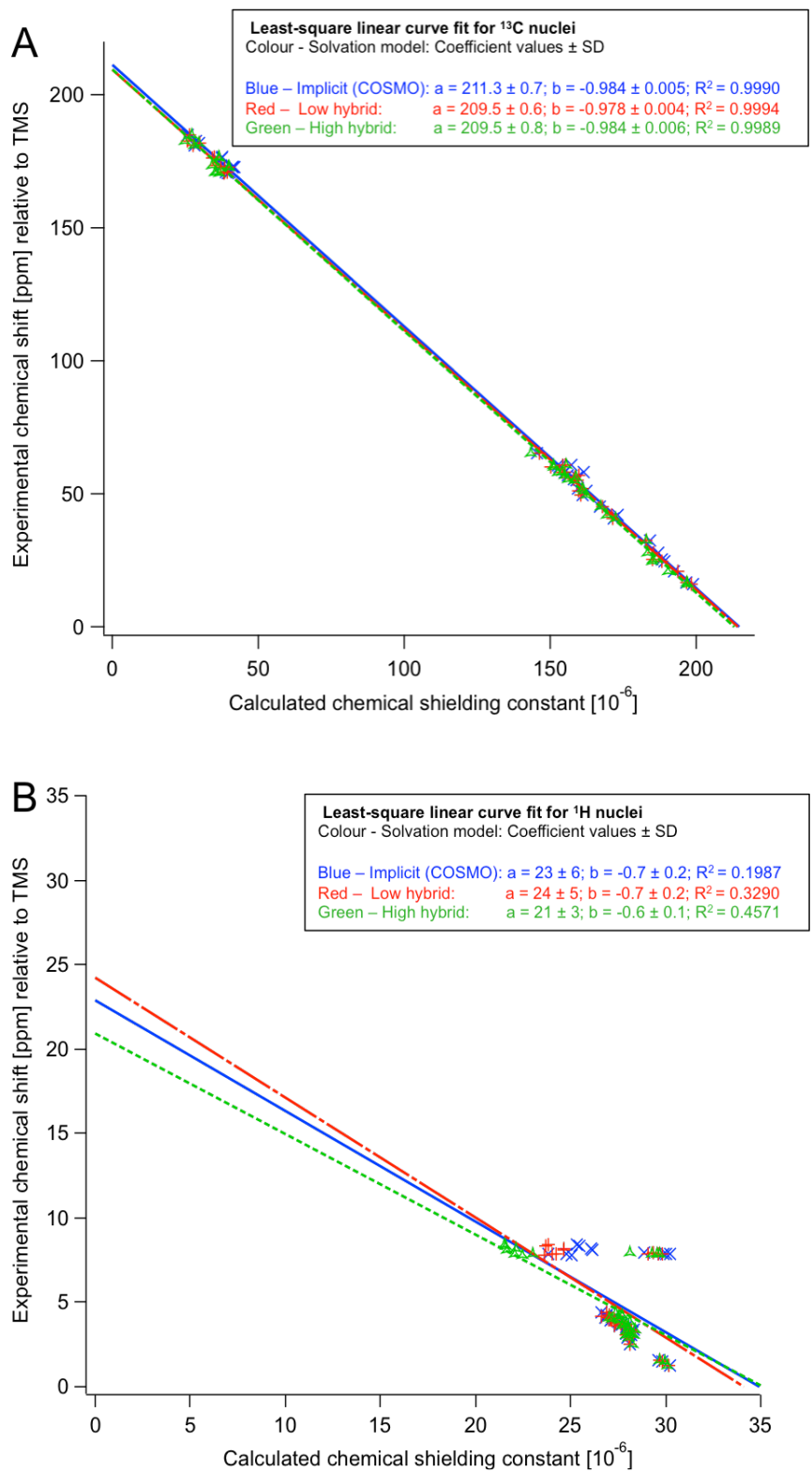

*Fig. 5: Correlation between nuclear shielding values calculated in implicit, low or high hybrid solvation and experimental chemical shift values} for $^{13}C$ (A) and $^{1}H$ (B) nuclei. For both nuclei types calculations were performed with implicit solvation (COSMO) (blue, continuous line), low hybrid solvation (red, intermitted line) and high hybrid solvation (green, dotted line).*

Protons bound to carbon atoms ($H_C$) were found to show the expected linear correlation. In contrast, protons bound to nitrogen ($H_N$) did not show the expected correlation due to the outliers with shielding values of around $30 \times 10^{-6}$ (Fig. 6). These outliers were identified to originate from calculated shielding constants of protons in amino groups of the deprotonated amino acid states. A closer look at the optimised geometries revealed that in all (fully) deprotonated states the nitrogen of the amine group formed a hydrogen bond with a water molecule. In all zwitterionic and protonated forms, however, the hydrogen atoms of this amino group formed hydrogen bonds with the surrounding water. These hydrogen bonds between amino group $H_N$ and oxygen of water molecules clearly improved the correlation with experimental data by "shifting" the calculated values to the left compared to implicit solvation with no explicit water molecules (= no H-bonds). Indeed, the successive shift from $H_N$ values in blue, red and green to the left can be seen in Fig. 5B.

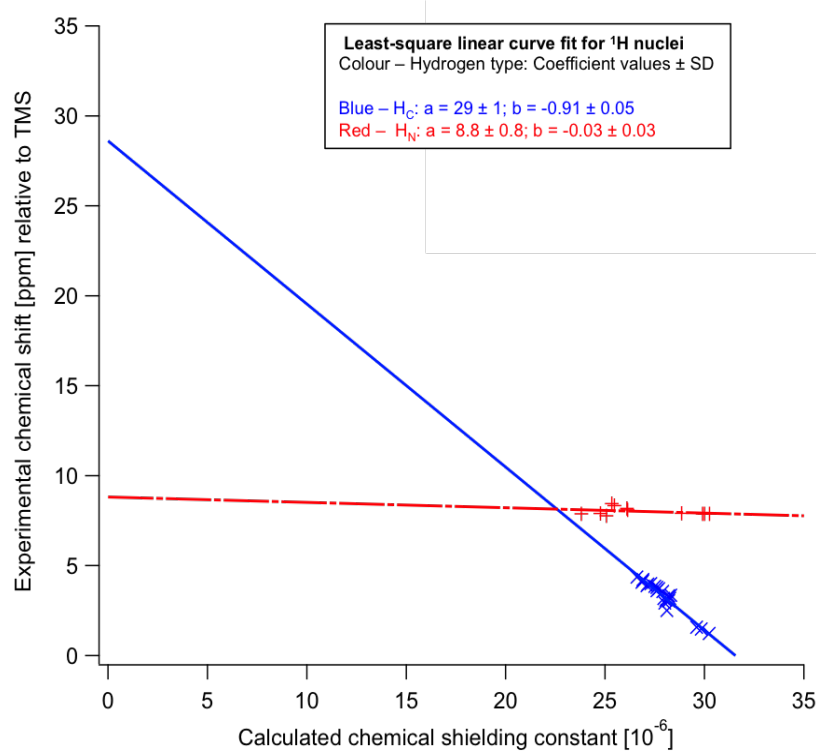

*Fig. 6: Comparison between correlation of $^1H$ nuclei attached to carbon (blue, continuous line) or nitrogen (red, intermitted line) calculated with COSMO solvent.*

In order to investigate why deprotonated amino acid states wouldn't form the same hydrogen bonds in more detail, we chose the glycine anion as an example. We found that the partial charge at the amino group $H_N$ in the anion was significantly lower than in the zwitterionic and protonated state (see Supplementary Material, Fig. S2). This could be caused by the overall negative charge of the molecule due to the deprotonated carboxyl group. This seems to cause a Coloumb repulsion between the amino $H_N$ and the oxygen of water and so make the interaction of the $H_N$ with water less favourable. We were able to optimise the system at a $H_N$–water distance of 2.4 Å (see Supplementary Material, Fig. S3), which corresponds to a weak hydrogen bond[62] and yields a nuclear shielding constant of $29 \times 10^{-6}$. We also observed that when the $H_N$ of anionic glycine were constrained at a given distance to a hydrogen-bond acceptor, such as the oxygen atom of a water molecule, their nuclear shielding values are affected proportionally to the distance (Supplementary Material, Fig. S4). This has been previously also observed by Moon & Case[63]. Based on the fitted correlation curve in Fig. 5, we would expect $H_N$ shielding constants of around $25 \times 10^{-6}$, which would correspond to a distance of 1.7 - 1.8 Å. The DFT methodology chosen in this study was found to be not suitable to correctly calculate this hydrogen bond interaction in deprotonated protonation states because electrostatic forces seem to overpower the forces of the hydrogen bonds. D3 is known to have issues with charged systems and using the recently developed D4 dispersion interactions[64] in future studies will potentially address this. Here, we therefore excluded the values of amino $H_N$ of deprotonated states in the following.

Analysing the results for proton nuclear shielding constants after excluding the outliers revealed a much clearer picture as shown in Fig. 7. The addition of one water molecule per ionisable functional group in the low hybrid model improved the correlation to experimental data by 11.8% and the accuracy by $\pm 1.00$ ppm to $\pm 0.81$ ppm compared to implicit solvation. With the high hybrid model, correlation was improved by 13.0% and accuracy by $\pm 0.88$ ppm compared to implicit solvation. However, the low hybrid solvation model performed best with regards to accuracy. The use of hybrid solvation shows a clear improvement to the nuclear shielding calculations.

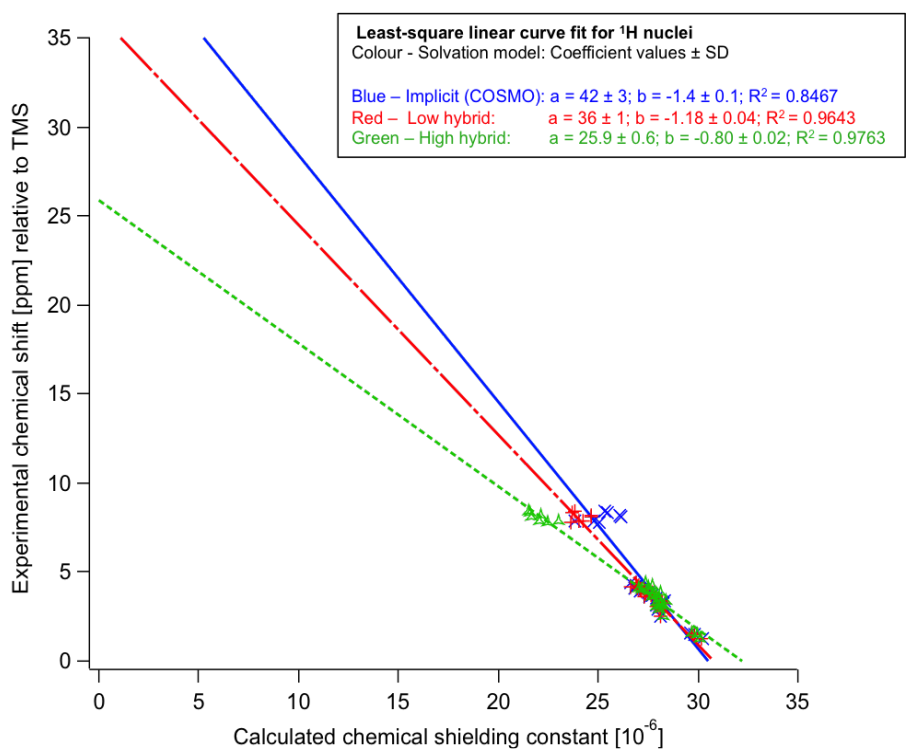

*Fig. 7: Correlation between calculated nuclear shielding values and experimental chemical shift values for $^1H$ nuclei excluding protons of the amino groups in anions. Calculations were performed with implicit solvation (COSMO) (blue, continuous line), low hybrid solvation (red, intermitted line) and high hybrid solvation (green, dotted line).*

## 3.4 Discussion

In this paper we illustrate the importance of solvation and the impact of different solvation models in the context of nuclear shielding calculations. It was established that the inclusion of solvent is crucial to obtain more accurate nuclear shielding predictions, irrespective of the implicit solvation model or program used. The addition of explicit solvent molecules to achieve a more accurate solvent representation is essential to represent the direct solute-solvent interactions like hydrogen bonds. It significantly improves calculated nuclear shielding values, especially those of protons. Using on average one water molecule per ionisable group in a hybrid approach was found to be sufficient.

**Performance of the hybrid models**

The low hybrid solvation model was found to perform best. It showed higher accuracies than the high hybrid solvation model in all cases and only minor differences in correlation. This contradicted our expectations because experimental and theoretical investigations on gas phase hydration show that for stabilisation of the zwitterionic form of non-polar amino acids, for example, 5 or more water molecules are required.[65,66,67,68,69] The worse performance of the high hybrid model seems to be caused by an overrepresentation of the solute-solvent interaction at the amino terminus. This significantly influences the linear fit and accuracy by affecting the slope as the amino $H_N$ nuclei are located at the top end of the graph. A similar observation of overestimation has also been reported by Cossi & Crescenzi for $^{17}O$ shieldings of small organic molecules when optimised in water clusters.[12] The averaged shielding values of the hydrogen atoms of the amino group significantly determine the slope of the regression and therefore the accuracy. Potentially the addition of three water molecules at the amino terminus in combination with the implicit solvation does not represent a realistic situation. Results of Panuszko *et al.* suggest an average number of 2.6 – 2.7 water molecules per amino group to form direct interactions in hydration.[66] This means that not all hydrogen atoms in the amino group are saturated by water molecules at all times. The lower number of water molecules used in the low hybrid model might therefore explain why it is performing better. For polar amino acids 1– 4 water molecules were found to be sufficient to stabilise the zwitterionic form in gas phase hydration.[65] In a wide range of protein crystal structures the amino acid:water ratio was determined as 0.4 to 2.7 depending on the polarity and solvent accessibility of the amino acid in a given protein conformation.[70] Therefore one water molecule per titratable group would present a suitable, more realistic and computationally less expensive approach for peptides, too.

**Method performance differs with nucleus type**

The performance of computational methods in comparison to the experimental data can be evaluated based on the slope of the linear regression between experimental and calculated data. This slope should be within a range of $-1.00 \pm 0.05$ in order to indicate only minimal systematic error.[9] The set of computational methods chosen here were found to perform very well for carbon nuclei as the regression slopes of all solvation models were $-1.00 \pm 0.02$. Systematic errors causing a change in the slope and therefore requiring significant scaling were not apparent. In contrast, for proton shieldings the slope values obtained for this test set deviated by up to 20% from the expected $-1.00$ when hybrid solvation models are used. This indicates the presence of a systematic error, even when the identified outliers are excluded. However, it has to be stressed that, in contrast to most other published studies, here we also included the nuclear shielding values of protons in amine groups ($H_N$). These protons are known to be very sensitive to molecular geometry and the formation of hydrogen bonds.[71,72,73] Therefore they provide lots of information and play a significant role in determining the regression parameters, as already discussed above. However, they have often been excluded in studies,[1] partly also due to experimental difficulties in their determination and their particular dependence on the solvent used.[74] In our study, the $H_N$ shieldings of protonated and zwitterionic amino acid forms were found to correlate reasonably well with the experimental data that could be obtained for 29 of 35 protonation states. Including the direct solute-solvent hydrogen-bond in the low hybrid model significantly improved the accuracy by decreasing the average shielding of the amino group protons. The inclusion of explicit water molecules was also reported to improve the agreement between calculated and experimental solution-phase NMR data of the amino $^1H$ chemical shift in guanine.[72] Exner et al. also documented that the inclusion of explicit solvent can significantly increase the chemical shift of $H_N$ by 2.34 ($\pm$ 1.2) ppm,[75] which is comparable to the changes observed in our study with 1.5 ($\pm$ 1.2) ppm.

**Reliability and accuracy in comparison to other approaches**

A chosen computational method always aims to strike a suitable balance between reliability and validity with the lowest computational costs possible. Our chosen method was found to provide reliable results for carbon and hydrogen nuclei including $H_N$, when the low hybrid approach is used, with overall very high correlation coefficients ($R^2$) of ≥95%. Besides its reliability for amino acids, it was further found to achieve good accuracies of carbon and proton nuclear shielding calculations in comparison to literature values in general. For carbon nuclei the accuracies reported in the literature range from $\pm$ 5 ppm to $\pm$ 1.53 ppm [76,60,59,77,1] depending on the level of theory and the nature of the dataset they are compared to. Best accuracies were reported for an approach using CCSD(T) with a large basis set and accounting for vibrational effects for a set of very small molecules in gas phase.[77] With DFT functionals the best accuracies for carbon nuclear shielding compared to experiment were reported by Frank et al. for the 32 amino acid long HA2 domain of the influenza virus glycoprotein hemagglutinin with $\pm$ 1.53 ppm (mPW1PW91/6-311G(d) with PCM and point charges in environment).[1] The same method yielded accuracies of $\pm$ 3.44 and $\pm$ 2.51 ppm for other parts of the same molecule.[1] These accuracies are similar to the values reported recently by Benassi, who tested a number of DFT functionals and noted WP04/DGTZVP amongst the best performing DFT functionals for a test set of organic molecules with an accuracy of $\pm$ 3.68 ppm.[41] The accuracy of $\pm$ 2.5 obtained for carbon nuclear shielding in our study (PBE0/aug-pc2) is therefore very good, albeit for a small data set of four amino acids. Hydrogen nuclear shieldings are reported with accuracies of between $\pm$ 0.86 ppm and $\pm$ 0.2 ppm in the literature.[1,76,78] The best accuracy obtained for protons most recently reported was $\pm$ 0.11 ppm, calculated with the WP04 functional (specific for chloroform solvation) and a 6-311+G(2d,p) basis set by Benassi for a test set of organic molecules with solvation in chloroform.[41] The PBE0 functional with the same basis set was found to perform almost equally well with an accuracy of $\pm$ 0.12 ppm in the same study.[41] This is a higher accuracy than we obtained in this study, but using our low hybrid solvation model we achieved an accuracy in the range reported by literature. This is very good considering the fact that, in contrast to the cited studies, our test set contains different protonation states and therefore differently charged molecules. It further includes not only the shieldings of protons attached to carbon but also to nitrogen atoms, which are usually excluded.

**The importance of test set nature and a consistent experimental data set**

The quality of the experimental data used for comparison is besides the computational methods the most crucial factor to obtain a realistic estimation of accuracy. It can profoundly affect the determined accuracy in two ways. Firstly, the type and diversity of molecules and molecular protonation states included in the test set determines how representative the obtained accuracy is for any other given molecule. The accuracy obtained for a test set with small organic, mostly rigid molecules in gas phase is unlikely comparable to the accuracy obtainable for nuclear shieldings of a test set with more complex, flexible biomolecules in aqueous solution. This is illustrated by the range of accuracies obtained with similar computational methods for different test sets (e.g. comparing the benchmarking of Flaig *et al.*[76] and Benassi[41]). Secondly, the experimental data itself can contain variabilities in the chemical shift values due to differences in NMR equipment, temperature, solvent, concentration or reference compound and referencing methods.[74,79,80] However, there are very few consistent experimental data sets with the same or at least comparable physicochemical factors, such as temperature and concentration. The accuracy values reported in the literature are always only representative of their particular test set, making a direct comparison difficult. In many cases the test sets are dominated by small organic molecules, for example in the studies of Rablen[78] or Auer *et al.*[77]. Furthermore, most computational results are compared to experimental data obtained by different working groups at different conditions. This is inevitably the case for large test sets containing a large number of compounds (e.g. in Benassi's study[41]). Our consistent, high quality experimental data set with an accuracy of $\pm$ 0.001 ppm for proton and $\pm$ 0.002 ppm for carbon shifts obtained with constant temperatures and consistent concentrations across samples in the same solvent with the same reference compound minimises the influence of variations in the experimental data on the analysis of shielding accuracies. It allows us to obtain a very good picture of the accuracy of the calculated shielding values alone. The results for this test set can only be compared indirectly to other approaches as biomolecules have only recently been considered in NMR test sets and amino acids are often not included. Our methodological approach and the obtained accuracies for a test set focussing on amino acid protonation states therefore provides a meaningful benchmark of particular importance for studies on amino acids, peptides and proteins.

**The indirect solvent effect and conformational sampling**

Apart from the direct and obvious effects of hydrogen bond formation and changes to the electrostatic environment, explicit solvent can also have an indirect effect that can influence the calculated shieldings: conformational change. As Cossi & Crescenzi showed very clearly in their study of small organic molecules in aqueous solution, the direct solvent effect can be well separated from an indirect effect caused by the water molecules altering the molecular conformation of the solute with approximately 10% the size of the direct solvent effect.[12] However, Monajjemi *et al.* reported in their study that the solvent-induced shielding variation is more likely influenced by the intensity of the solvent reaction field than the molecular geometry induced by the solvent.[81] Degtyarenko *et al.* concluded from a molecular dynamics (MD) study of L-alanine in aqueous solution that the first hydration shell of amino acids is localised around the carboxylate and ammonium functional groups.[82] This shell is highly ordered and quite rigid but the participating water molecules were found to constantly exchange.[82] This would agree with our findings for the hybrid models. In order to represent the varying solute-water interactions better, conformational averaging would be a good potential technique. Its importance has been increasingly highlighted over the past years.[83,84,85] Results of Kwan *et al.* and Exner *et al.* suggest that averaging over MD snapshots for which the chemical shift values are calculated individually yields a significant improvement of accuracy.[83,75] The use of only one optimised lowest energy conformer for each protonation state in our study will therefore have to be compared to results obtained with methods using MD and conformational averaging (for example as described by Dračínský *et al.*[85]) in the future. This could also help to resolve the problems observed for $H_N$ nuclei in deprotonated amino acid states.

# 4. Conclusion

This study highlights that the inclusion of solvent is crucial to obtain more accurate nuclear shielding predictions, irrespective of the implicit solvation model or program used. The addition of explicit solvent molecules to achieve a more realistic representation is essential to account for the direct solute-solvent interactions such as hydrogen bonds. It significantly improves calculated nuclear shielding values, especially those of protons. Using on average one water molecule per ionisable group in a hybrid approach was found to be sufficient to achieve a good accuracy.

Nevertheless, the use of a single optimised lowest energy conformer for each protonation state in our study does not allow any conformational averaging. Future work should compare this approach to results obtained with methods using MD and conformational averaging as this could also help resolve the issues observed for $H_N$ nuclei in deprotonated amino acid states.

In contrast to other studies, our test set contains different protonation states and therefore differently charged molecules. It further includes not only the shieldings of protons attached to carbon but also to nitrogen atoms, which are usually excluded. Moreover, it provides exceptionally consistent experimental data set for comparison (to be released later). Our study suggests that using one explicit water molecule per titratable group would present a suitable, realistic and computationally inexpensive approach to determine NMR shielding for peptides.

# Acknowledgments

We acknowledge the Viper High Performance Computing facility of the University of Hull and its support team as well as funding for CCR's PhD scholarship by the University of Hull.

# Supplementary Material

*Table T1: Comparison of aug-pc-2 vs. aug-pcS-2*

| Nucleus | aug-pc2 | | aug-pcS2 | |
|---|---|---|---|---|
| | Accuracy | $R^2$ | Accuracy | $R^2$ |
| $^{13}C$ | ± 2.9 | 1.0 | ± 4.0 | 1.0 |
| $^1H$ | ± 1.1 | 0.2 | ± 2.2 | 0.16 |

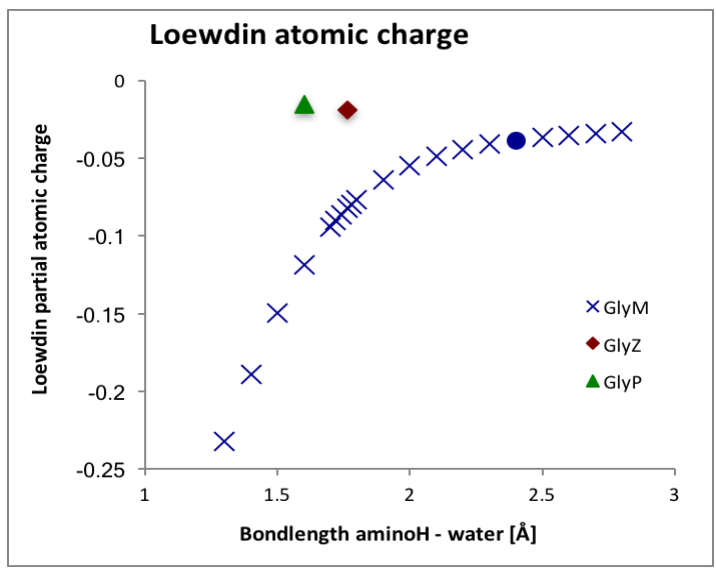

*Fig. S2: Partial charge of the $H_N$ of the glycine anion (blue) with varying bond length. Blue dot indicates energy lowest conformer. Partial charge for $H_N$ of the optimised glycine zwitterion and protonated form are also shown (red and green, respectively).*

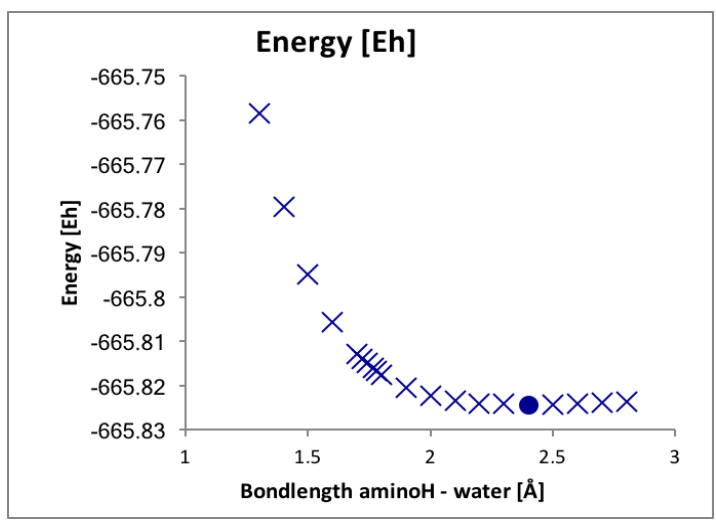

*Fig. S3: Energy of the glycine anion (blue) with varying bond length. Blue dot indicates energy lowest conformer.*

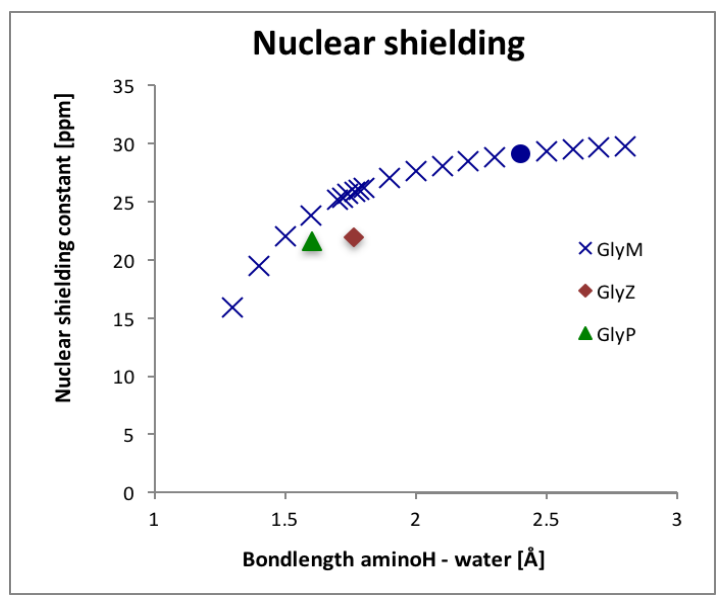

*Fig. S4: Nuclear shielding of the $H_N$ of the glycine anion (blue) with varying bond length. Blue dot indicates energy lowest conformer. Partial charge for $H_N$ of the optimised glycine zwitterion and protonated form are also shown (red and green, respectively).*